\documentclass[conference]{IEEEtran}

\usepackage{graphicx}
\usepackage{subfigure}
\usepackage{amsmath}
\usepackage{amsfonts}
\usepackage{amssymb}
\usepackage{latexsym}
\usepackage{moreverb}
\usepackage{booktabs}
\usepackage{times}
\usepackage{indentfirst}
\usepackage{algorithmic}
\usepackage{algorithm}
\usepackage[table]{xcolor}  % for table with color
\usepackage{multirow}       % for table

\usepackage{array}
\usepackage{balance}
\usepackage{comment}
\usepackage{fancyhdr}

\graphicspath{{eps/}}

\usepackage[left=.6in,right=.6in,top=.6in,bottom=.6in]{geometry}

\usepackage{mdwmath}
\usepackage{mdwtab}
\usepackage{eqparbox}
\usepackage{url}

\newcommand{\Fig}[1]{Fig.~\ref{#1}}

\fancypagestyle{IEEE_Green_open_access_footer}{%
\vspace{-1cm}
  \fancyhf{}
  
  \fancyfoot[C]{\footnotesize \copyright2014 IEEE. Personal use of this material is permitted. Permission from IEEE must be obtained for all other uses, in any current or future media, including reprinting/republishing this material for advertising or promotional purposes, creating new collective works, for resale or redistribution to servers or lists, or reuse of any copyrighted component of this work in other works. DOI: 10.1109/MASS.2014.32}
}

\begin{document}

\title{WiFiScout: A Crowdsensing WiFi Advisory System with Gamification-based Incentive}

\author{Fang-Jing~Wu and Tie Luo\\
Institute for Infocomm Research, A*STAR, Singapore\\
Email:wufj@i2r.a-star.edu.sg, luot@i2r.a-star.edu.sg \\
}

\maketitle
\thispagestyle{IEEE_Green_open_access_footer}

\begin{abstract}
As mobile crowdsensing techniques are steering many
smart-city applications, an incentive scheme that motivates
the crowd to actively participate becomes a key to the success
of such city-scale applications.
This paper presents a crowdsensing WiFi advisory system called
\emph{WiFiScout}, which helps smartphone users to find
good quality WiFi hotspots. The quality information is defined
in terms of user experience and hence the system requires users to contribute
information of their experience with WiFi hotspots.
To motivate people to contribute such information, we design and implement
a gamification-based incentive scheme in WiFiScout. It allows
a user to ``conquer WiFi territories'' by becoming the top
contributor for WiFi hotspots at different locations. The contribution
is based on the diversity and amount of data a user
submits, for which he will be rewarded accordingly.
WiFiScout has been implemented on Android
and it facilitates the collection of city-wide WiFi advisory information
provided by real users according to their actual experience.
\end{abstract}

%===== ACM classification sections (Categories, General Terms and keywords) ======
%\category{C.3}{SPECIAL-PURPOSE AND APPLICATION-BASED
%SYSTEMS}[Real-time and embedded systems]
%
%\terms{Algorithms}
%

\section{Introduction}
Mobile crowdsensing techniques have inspired many real-world
smart-city applications such as traffic navigation systems \cite{Waze} and
urban noise mapping systems \cite{Rana2010_Ear-Phone}. Some
other systems provide coarse-grained information
of WiFi hotspots \cite{WeFi} or cellular network coverage \cite{OpenSignal}.

However, a big challenge in crowdsensing applications is to have an
incentive scheme that motivates people to contribute sensing data.
This work presents a crowdsensing WiFi advisory
system called \emph{WiFiScout}, that incorporates a gamification-based
incentive scheme to reward users who contribute the most useful data
based on the diversity and amount of the contributed data.
WiFiScout supports three advisory modes: (1) offline search,
(2) online review, and (3) gamification-based WiFi map. The offline search
mode allows a user to search for a list of available WiFi access points
around a queried region, even though he is not near the region.
The online review mode allows a user who already connects to a WiFi
access point to submit a review about his experience on that WiFi
access point through his smartphone. The gamification-based WiFi
map displays all the WiFi access points on a city map, but unlike other
similar applications, each access point is represented by a user
who has contributed the most useful information to it.
Compared to existing systems, WiFiScout has the following unique features.
First, it incorporates a gamification-based incentive to motivate
people to contribute high-quality data \cite{qcs13}, in which the reward system
considers not only the amount but also the diversity of
the contributed data. Second, WiFiScout collects real user
experience---in terms of WiFi network quality and
tagged by semantic places---as well as device-measured metrics such
as signal strength and link speed, in order to provide more
comprehensive WiFi-related information.

\begin{figure}
\centering
\includegraphics[width=0.95\columnwidth]{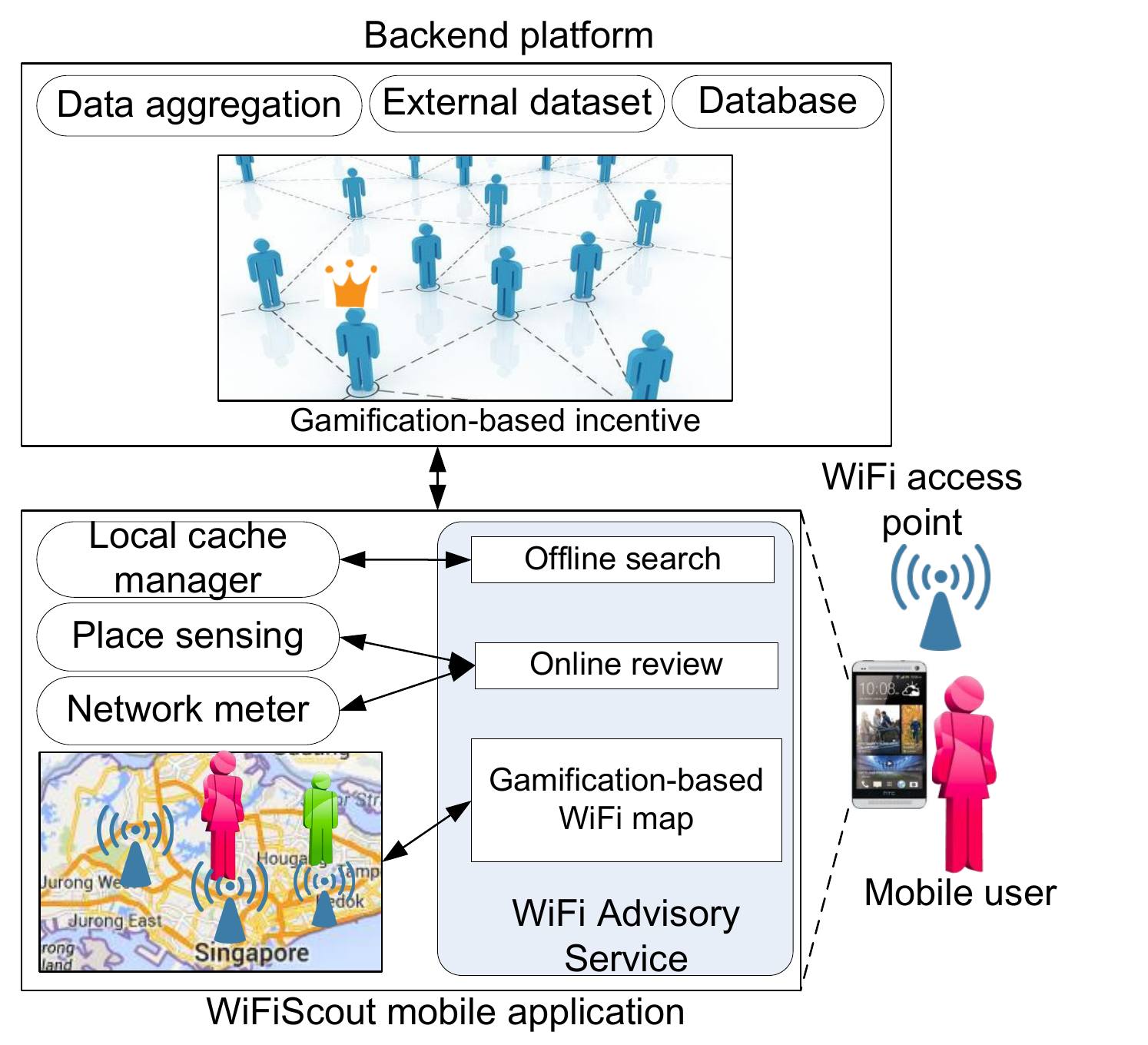}
\caption{The system architecture of WiFiScout.}
\label{Fig:SysArchitecture}
\end{figure}

\begin{figure*}
\centering
\includegraphics[width=1.9\columnwidth]{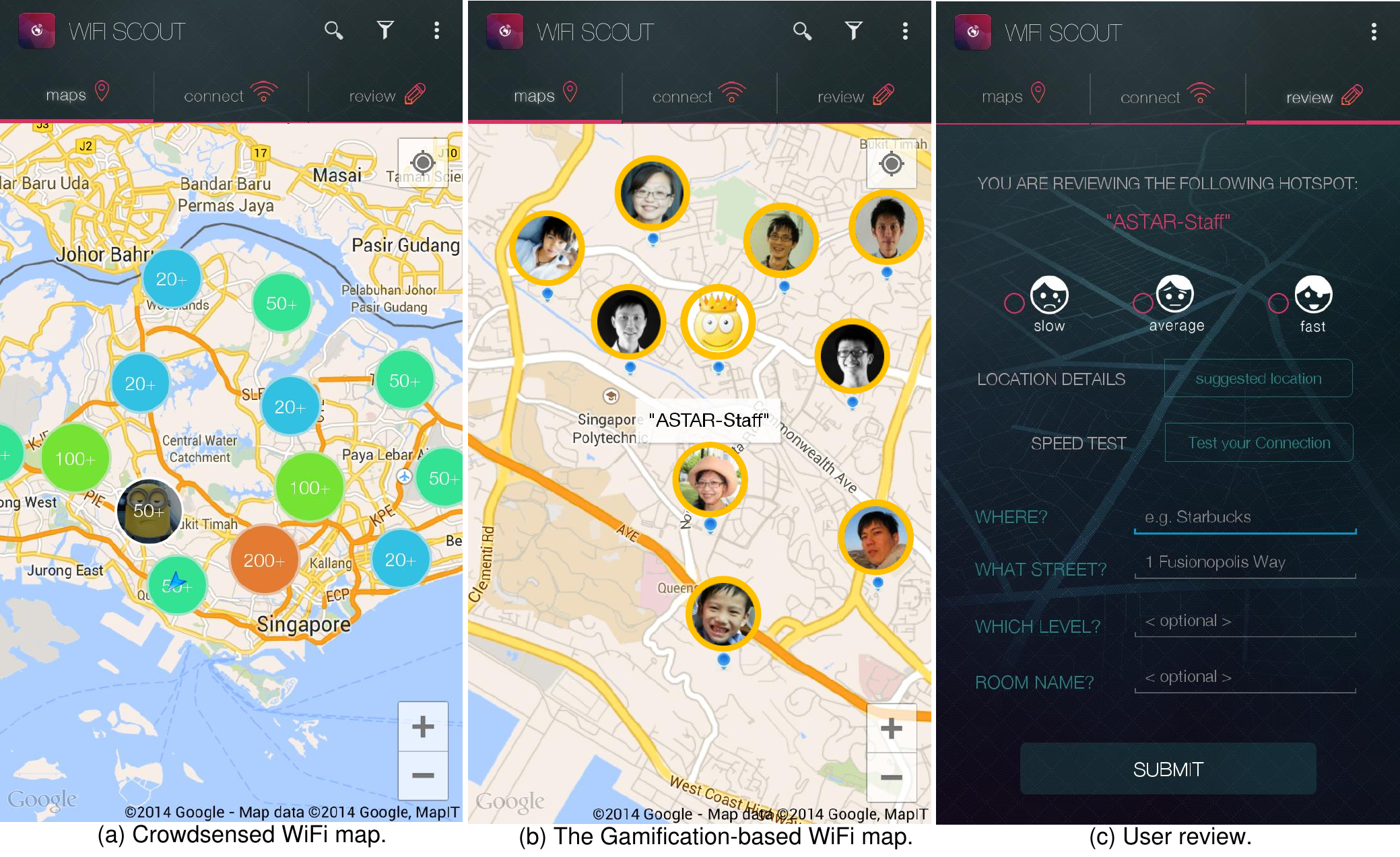}
\caption{Screenshots of the WiFiScout mobile application.}
\label{Fig:DemoUI}
\end{figure*}

\section{System Design}

\subsection{System Architecture}
See \Fig{Fig:SysArchitecture}. The
front-end WiFi advisory service cooperates with the
gamification-based incentive to collect diverse and high-quality
data. The back-end platform manages the collected data and
evaluates users' contributions through four major components:
database, external dataset, data aggregation, and
gamification-based incentive. The database stores the raw
sensing data and processed data. Not only our own dataset but also
an external dataset from a government agency are imported into our
system to enhance the WiFi advisory service. The component of data
aggregation clusters collected data based on the spatial
correlation among them. The gamification-based incentive evaluates
each user's contributions based on the data diversity submitted by
the user and let all the users compete for rewards.

\subsection{Front-end Design}
WiFiScout consists of four major components:
(1) WiFi advisory service, (2) local cache manager,
(3) place sensing, (4) network meter. The WiFi advisory service
guides people to find out ``better" WiFi access point,
collects user reviews, and provides users with a
gamification interface to win rewards. The local cache
manager maintains a lightweight database on each smartphone
to support offline search.
The place sensing captures the semantic place information
associated with a WiFi access point, represented by a fine-grained
location including street address, floor information, and room
information, in order to facilitate a user to find the WiFi access point
easily. The network meter measures the connectivity quality of a
WiFi access point including the signal strength,
link speed, uploading speed and downloading speed, associated
with its BSSID and SSID.

\subsection{Reward Evaluation}
The gamification-based incentive scheme evaluates user
contributions according to the following rules:
\begin{itemize}
    \item When a new user initially joins the system, the user
    will have $R_s$ starting points.
    \item When a user $u$ connects to a WiFi access point, say $AP$,
    and submits a user review for $AP$, he will be rewarded points of:
    \begin{itemize}
        \item $R$, if $u$ did not contribute any review for $AP$ in the past;
        \item $\frac{R}{2}$, if $\tau(AP,u)\ge T$, where $\tau(AP,u)$ is the time interval
        between this review and the previous review for $AP$ submitted
        by $u$, and $T$ is a predefined threshold; or
    \item 0, if $\tau(AP,u)<T$.
    \end{itemize}
\end{itemize}
By these rules, a user who reviews many different
WiFi access points with sufficient time interval in between
(so that there is less correlation), and for a long term,
will be considered by the system as a useful data contributor.
Thus, the gamification-based incentive will rank users
according to their reward points, and enable the top contributor
for each WiFi access point to ``own'' that access point.
As this process continues, such ownership will change according
to each user's contribution performance and, thereby, creating
a gamification enthronement in which users compete with one another
to ``claim'' territories by placing their avatar pictures on as many locations
as possible in a city map.

\section{Implementation and Demonstration}
In the current implementation of WiFiScout, we choose the threshold
$T$ to be 6 hours. \Fig{Fig:DemoUI}(a) is the crowdsensed
WiFi map where nearby hotspots are clustered together, with
the number indicating the cluster size. \Fig{Fig:DemoUI} (b) is
the gamification-based WiFi map where each avatar picture represents
the top contributor for that particular WiFi access point. \Fig{Fig:DemoUI} (c)
is the interface crowdsourcing for the user reviews on their WiFi experience.

The next implementation will enhance both incentive and trustworthiness
aspects by incorporating a social-economic scheme called SEW \cite{sew14}.
\section*{Acknowledgment}
The authors would like to thank Wendy Ng, Huiwen Zhong,
Johnny Zhou, Jason Cheah, La Thanh Tam and Mingding Han for their
help in developing the system.

%%%%%%%%%%%%%%%%%%%%%%%%    References  %%%%%%%%%%%%%%%%%%%%%%%%%%%%%%%
\bibliographystyle{IEEEtran}                                          %
%\bibliography{wifiscout}                                              %
%%%%%%%%%%%%%%%%%%%%%%%%%%%%%%%%%%%%%%%%%%%%%%%%%%%%%%%%%%%%%%%%%%%%%%%


\end{document}